\documentclass[floatfix,nofootinbib,twocolumn,showpacs,superscriptaddress, reprintnumbers,amssymb,letterpaper,amsmath,pra,amsfonts]{revtex4-2}

\usepackage{graphicx}
\usepackage{dcolumn}
\usepackage{bm}
\usepackage{xcolor}
\usepackage[colorlinks=true,allcolors=blue]{hyperref}
\usepackage{amsmath}
\usepackage{lineno}
\usepackage{comment}
\usepackage{tikz}
\usepackage{url}
\usepackage{cleveref}
\usetikzlibrary{shapes.geometric, arrows}

\begin{document}
\title{Focusing of Relativistic Electron Beams With Permanent Magnetic Solenoid}

\author{T. Xu}
\email{txu@slac.stanford.edu}
\affiliation{SLAC National Accelerator Laboratory, Menlo Park, CA 94025, USA} 
\author{C. J. R. Duncan}
\affiliation{SLAC National Accelerator Laboratory, Menlo Park, CA 94025, USA} 
\author{P. Denham}
\affiliation{Department of Physics and Astronomy, UCLA, Los Angeles, California 90095, USA}
\author{B. H. Schaap}
\affiliation{Department of Physics and Astronomy, UCLA, Los Angeles, California 90095, USA}
\author{A. Kulkarni}
\affiliation{Department of Physics and Astronomy, UCLA, Los Angeles, California 90095, USA}
\author{D. Garcia}
\affiliation{Department of Physics and Astronomy, UCLA, Los Angeles, California 90095, USA}
\author{S. D. Anderson}
\affiliation{SLAC National Accelerator Laboratory, Menlo Park, CA 94025, USA} 
\author{P. Musumeci}
\affiliation{Department of Physics and Astronomy, UCLA, Los Angeles, California 90095, USA}
\author{R. J. England}
\affiliation{SLAC National Accelerator Laboratory, Menlo Park, CA 94025, USA} 


\date{\today}

\begin{abstract}
Achieving strong focusing of MeV electron beams is a critical requirement for advanced beam applications such as compact laboratory X-ray sources, high gradient accelerators, and ultrafast electron scattering instrumentation. To address these needs, a compact radially magnetized permanent magnetic solenoid (PMS) has been designed, fabricated, and tested. The solenoid provides a compact and inexpensive solution for delivering high axial magnetic fields (1 Tesla) to focus MeV electron beams. Field characterization of the solenoid demonstrates excellent agreement with analytical models, validating the PMS design. The electron beam test employs a high-brightness photoinjector to study the focusing properties of the PMS. The results indicate a focal length of less than 10 cm and a significant reduction in beam size with small spherical aberrations. Two application cases are evaluated: angular magnification in ultrafast electron diffraction setups and strong focusing for Compton scattering or other microfocus uses.
\end{abstract}

\maketitle


\section{Introduction \label{sec:intro}}


The generation of relativistic electron beams with MeV-scale energies from rf photoinjectors has revolutionized ultrafast science and advanced photon sources~\cite{Musumeci2018review}. Characterized by sub-picosecond temporal duration and sub-micron transverse emittance, these beams are indispensable across diverse domains including ultrafast electron diffraction (UED)~\cite{Filippetto2022review}, inverse Compton X-ray sources~\cite{graves2014}, compact high-frequency accelerators~\cite{england2014, zhang2018segmented}, and MeV-scale electron microscopy and radiography~\cite{rkli2014, zhou2019demonstration}. In these applications, achieving strong focusing of relativistic electron beams is critical---whether to resolve atomic-scale structural dynamics through angular magnification in UED~\cite{denham_high_2024}, enhance photon production in compact light sources~\cite{lim_adjustable_2005}, or enable precise beam manipulation and injection in advanced acceleration schemes. However, the relativistic inertia of MeV electrons drastically reduces their sensitivity to transverse focusing fields, posing stringent demands on magnetic optics systems.


Conventional electromagnetic solenoids, though capable of delivering the required focusing, have several practical limitations such as significant power consumption, active cooling requirements, and a generally larger footprint---all factors that hinder their integration into space-constrained beamlines or experimental setups. Permanent magnets, which leverage rare-earth magnetic materials such as NdFeB, have emerged as a compelling alternative, offering high magnetic fields, passive operation, and compact geometries. Permanent magnetic quadrupoles, for instance, can access very high field gradients (exceeding 500 T/m) in millimeter-scale apertures culminating in compound lenses with focal lengths as short as a few cms~\cite{cesar_demonstration_2016,ghaith2019permanent,wan2018design}. Achieving stigmatic focusing (equal focal lengths in $x$ and $y$) with permanent magnet quadrupoles typically requires using a triplet configuration, with precise control over the longitudinal spacings between quadrupoles to near $\mu$m precision. 
Like quadrupole triplets, permanent magnetic solenoids (PMS) offer passive and compact focusing for MeV electron beams, but further provide axisymmetric focusing and greater experimental simplicity.~\cite{hoff2012wide,gehrke_pms_2013,hachmann_design_2016}. PMS designs can be classified into two categories based on magnetization orientation: axially magnetized (AM-PMS) and radially magnetized (RM-PMS) configurations. Previous studies have shown that dual-ring RM-PMS designs achieve higher peak on-axis fields compared to axially magnetized designs under the same size and weight constraints, making them particularly suitable for applications demanding high field strengths in limited spatial footprints. Pairing two PMS together can be used to tune the focusing strength, cancel multipole moments and reduce aberrations. 

In this paper, we present the design, field characterization, and electron beam test of a dual-ring RM-PMS optimized for UED applications. The PMS features a 1.2~cm bore diameter to accommodate relatively large beam sizes while achieving a peak on-axis field of 1~T. Comprehensive simulations and Hall probe field mapping confirm excellent agreement between modeled and measured field profiles. The RM-PMS design was based on a beam energy of 4 MeV typical for MeV-UED experiments. However, to evaluate focusing performance, we conducted electron beam experiments using a 7.1 MeV high-brightness photoinjector beamline, demonstrating a 8.4 cm focal length in agreement with the predictions. Notably, achieving a similar result with an electromagnetic solenoid would have likely required superconducting technology \cite{lefranc1982superconducting} with multiple orders of magnitude increase in size and cost for the electron lens. These results establish the PMS as a robust, cost-effective, cooling-free solution for strong focusing of relativistic beams, directly addressing the needs of many applications requiring compact, high-field focusing for MeV high brightness beams. After revisiting some design considerations, we present the magnetic measurements of the first prototype RM-PMS, followed by the presentation of the results of the relativistic beam tests. We then conclude discussing a couple of specific applications of a compact PMS including improving the angular magnification of the diffraction pattern in a UED setup, and the generation of tight beam waists for inverse Compton scattering sources~\cite{graves2014,Deitrick2018}, and other instances where sub-micron beam spot sizes are desired \cite{ji2019ultrafast, kulkarni24progress}.

\section{Radially magnetized PMS rings}

\subsection{Design and fabrication}\label{subsec:design}

\begin{figure}[t]
\centering
\includegraphics[width=0.92\columnwidth]{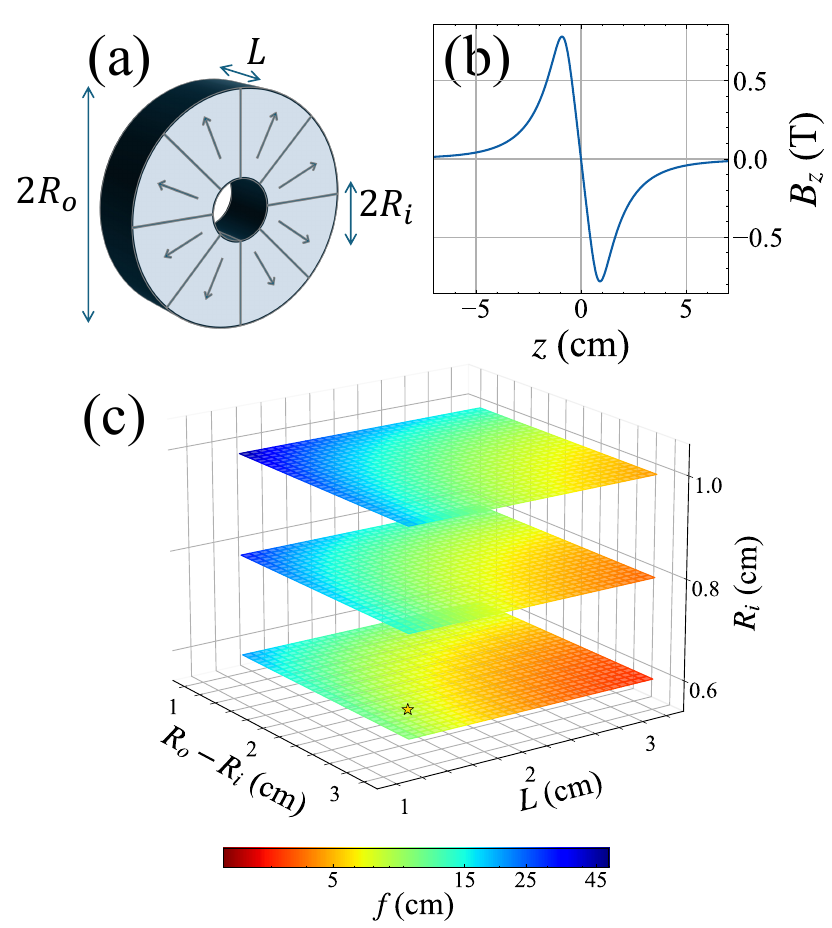}
\caption{(a) Cartoon design of a single RM-PMS ring. The arrows of in the segments denote the direction of magnetization. (b) On-axis field profile of a single RM-PMS ring with $R_{i}$=0.6~cm, $R_{o}$=3.5~cm, $L$=1.6~cm. (c) Focal lengths $f$ of single RM-PMS rings ($B_{r}=1.32$~T) as a function of wall thickness ($R_{o}- R_{i}$) and length $L$ for inner radii $R_{i}$=0.6,~0.8,~1~cm (electron beam kinetic energy 4~MeV). Different colors represent the corresponding focal lengths. The star in the plot denotes the final design parameters of the single RM-PMS ring with $R_{i}$=0.6~cm, $R_{o}-R_{i}$=2.9~cm, $L$=1.6~cm.}
\label{fig:pms_single_ring_f}
\end{figure}

The on-axis magnetic field of a single RM-PMS ring with inner radius $R_{i}$, outer radius $R_{o}$ and length $L$, as depicted in Figure~\ref{fig:pms_single_ring_f}(a), can be calculated using a surface current model~\cite{peng_axial_2004} and is given by
\begin{equation}
    B_{z,\mathrm{RM-PMS}} =  B_{z, \mathrm{disk}} (z + L/2) - B_{z, \mathrm{disk}} (z - L/2),
\label{eq:Bz_single_pms}
\end{equation}
where $ B_{z, \mathrm{disk}}$ is the on-axis field of a thin current disk and can be expressed as,
\begin{equation}
\begin{aligned} B _ {z,\text{disk}} ( z ) = \frac { B _ { r } } { 2 } \left( \ln \frac { \sqrt { z ^ { 2 } + R _ { o } ^ { 2 } } + R _ { o } } { \sqrt { z ^ { 2 } + R _ { i } ^ { 2 } } + R _ { i } } \right. & - \frac { R _ { o } } { \sqrt { z ^ { 2 } + R _ { o } ^ { 2 } } } \\ & \left. + \frac { R _ { i } } { \sqrt { z ^ { 2 } + R _ { i } ^ { 2 } } } \right),
\end{aligned}
\label{eq:Bz_disk}
\end{equation}
where $B_{r}$ is the remanence of the magnet. Equation~(\ref{eq:Bz_single_pms}) describes the field of a radially outward-magnetized PMS, and the field reverses sign for a radially inward-magnetized PMS. Figure~\ref{fig:pms_single_ring_f} (b) presents the field profile of a single RM-PMS ring (magnetized outward) calculated with Eq.~(\ref{eq:Bz_single_pms}) with $B_{r}$=1.32~T, $R_{i}$=0.6~cm, $R_{o}$=3.5~cm, $L$=1.6~cm. A notable property of RM-PMS, evident from Eq.~(\ref{eq:Bz_single_pms}) and Fig.~\ref{fig:pms_single_ring_f}(b), is that the integral of the on-axis field over the longitudinal axis vanishes, which implies RM-PMS produces no net Larmor rotation. 

We calculate the focal lengths of single RM-PMS rings of different wall thickness ($R_{o}-R_{i}$), length, and inner radii using the well-known formula $\frac{1}{f} = \left( \frac{e}{2 p_z} \right)^2 \int_{-\infty}^{\infty} B_z^2 \, dz$, where $e$ and $p_{z}$ are the electron charge and longitudinal momentum. The results for a 4~MeV kinetic energy beam are presented in Figure~\ref{fig:pms_single_ring_f}. In general, achieving shorter focal lengths requires using rings with thicker walls, longer lengths, or smaller inner radii, as these adjustments enhance the on-axis field. 

\begin{figure}[b]
\centering
\includegraphics[width=0.95\columnwidth]{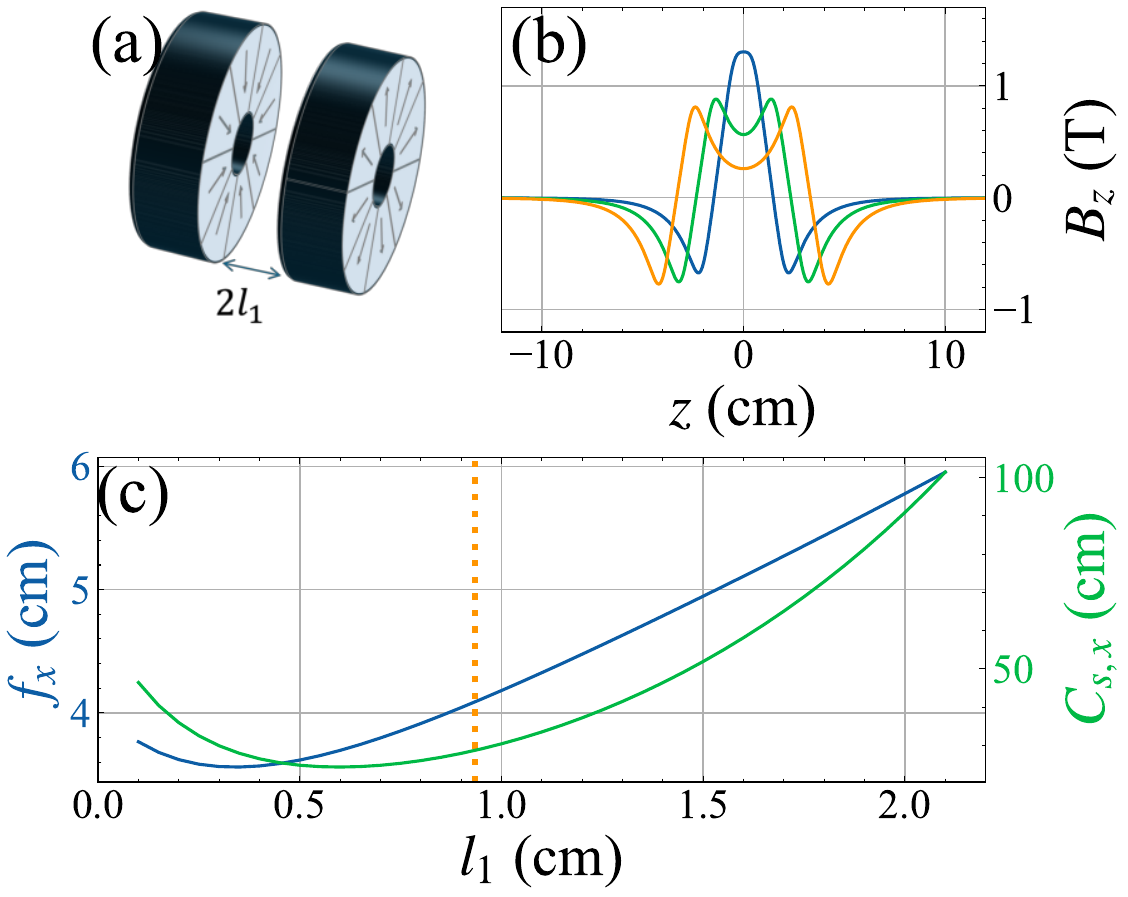}
\caption{(a) Schematic of dual RM-PMS ring arrangement. (b) On-axis field profiles of dual RM-PMS rings with $B_{r}=1.32$~T, $R_{i}=0.6$~cm, $R_{o}=3.5$~cm, $L=1.6$~cm, and different values of $l_{1}$. The blue, green, orange traces correspond to $l_{1}$=0.5, 1.5, 2.5 cm respectively. (c) Evolution of focal length (blue trace, left axis) and spherical aberration coefficient (green trace, right axis) for different values of $l_{1}$ at 4~MeV kinetic energy. The orange dotted line denotes the final design choice of $l_{1}$=0.935~cm.}
\label{fig:pms_design}
\end{figure}

The on-axis field can be further enhanced by combining two concentric RM-PMS rings with opposite radial magnetization directions (inward and outward) to form a dual RM-PMS ring. Following the notation in~\cite{gehrke_pms_2013} and using 2$l_{1}$ to denote the distances between the inner faces [see Figure~\ref{fig:pms_design}(a)], the on-axis field of the dual RM-PMS ring configuration is given by:
\begin{equation}
\begin{aligned}    
B_{z,\mathrm{dual}} & =   B_{z, \mathrm{disk}} (z + l_{1}) + B_{z, \mathrm{disk}} (z - l_{1}) \\ &  - B_{z, \mathrm{disk}} (z + l_{1} + L) - B_{z, \mathrm{disk}} (z - l_{1} - L).
\end{aligned}
\label{eq:Bz_dual_pms}
\end{equation}

As depicted in Figure~\ref{fig:pms_design}(b), the field profiles of the dual RM-PMS rings can be adjusted by varying the separation distances between the two single rings. To illustrate the effect of varying $l_{1}$ on focusing strength and field aberrations of the dual RM-PMS rings, we consider a configuration composed of two single rings with dimensions $R_{i}=0.6$~cm, $R_{o}=3.5$~cm, $L=1.6$~cm and a remanence of $B_{r}=1.32$~T. We compute the transfer maps in {\sc cosy infinity}~\cite{berz1990} for beams with 4~MeV kinetic energy, using transport matrix notation where $R_{ij}$ and $U_{ijkl}$ denote first- ($6\times6$) and third-order ($6\times 6\times 6\times6$) transfer matrix elements respectively~\cite{brown1982}. The focal length of the dual ring is determined by $f_{x}=-1/R_{21}$, while the spherical aberration coefficient is calculated as $C_{s,x} = U_{1222}/R_{11}$. As shown in Figure~\ref{fig:pms_design}(c), the spherical aberration coefficient generally scales with the focal length, and an optimal separation distance exists that minimizes either the focal length or the spherical aberration coefficient.

\begin{table}[t]
    \centering
    \begin{tabular}{cccc} \hline \hline
        \textbf{Parameter} & \textbf{Symbol} & \textbf{Value} & \textbf{Unit}\\ \hline
        Remanence & $B_{r}$ & 1.32 & T \\ 
        Inner radius & $R_{i}$ & 0.6 & cm \\ 
        Outer radius & $R_{o}$ & 3.5 & cm \\ 
        Length of single ring & $L$ & 1.6 & cm \\ 
        Center to inner face & $l_{1}$ & 0.935 & cm \\ 
        Focal length (single ring, 4~MeV) & $f_{\mathrm{s,4MeV}}$ & 7.8 & cm \\ 
        Focal length (single ring, 7~MeV) & $f_{\mathrm{s,7MeV}}$ & 19.3 & cm \\ 
        Focal length (dual ring, 4~MeV) & $f_{\mathrm{d,4MeV}}$ & 4.2 & cm \\ 
        Focal length (dual ring, 7~MeV) & $f_{\mathrm{d,7MeV}}$ & 8.4 & cm \\ \hline \hline
    \end{tabular}
    \caption{Final design parameters of the RM-PMS ring.}
    \label{tab:pms_parameter}
\end{table}

The final design parameters of the RM-PMS ring are summarized in Table~\ref{tab:pms_parameter}. These parameters were selected to minimize the focal length at 4~MeV (approximately 4~cm) while also considering practical constraints such as holder dimensions and stage compatibility. Raytracing simulations of the optimized design yield focal lengths of 4.2~cm and 8.4~cm for beams with kinetic energies of 4~MeV and 7~MeV respectively and are consistent with $-1/R_{21}$ calculated with {\sc cosy infinity}.

\begin{figure}[t]
\centering
\includegraphics[width=0.85\columnwidth]{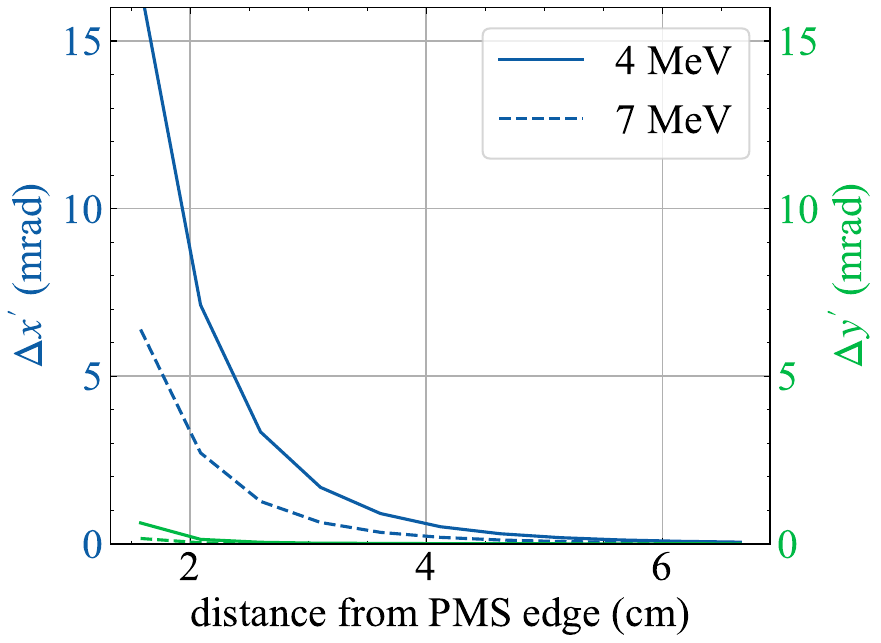}
\caption{Kick exerted on electron beam by PMS fringe fields when the magnet is moved off-axis horizontally. Blue traces denote the kick in $x$ (left axis) and green traces the kick in $y$ (right axis). }
\label{fig:pms_steering}
\end{figure}

\begin{figure*}[t] 
\centering
\includegraphics[width=1.65\columnwidth]{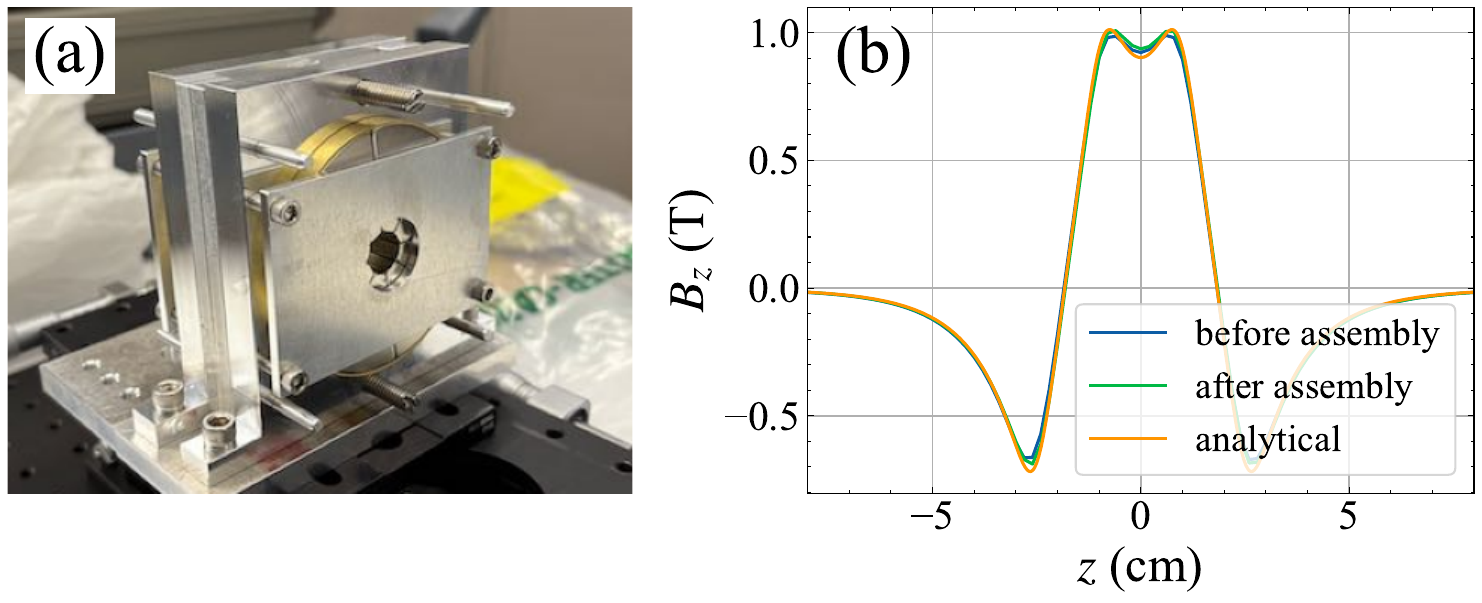}
\caption{(a) Fully assembled dual RM-PMS ring. (b) On-axis magnetic field of the dual RM-PMS ring comparing a superposition of the pre-assembly single-ring measured field profiles (blue), measured post-assembly dual-ring field (green), and analytical calculation (orange).}
\label{fig:pms_measurements}
\end{figure*}

\begin{figure}[b] 
\centering
\includegraphics[width=0.96\columnwidth]{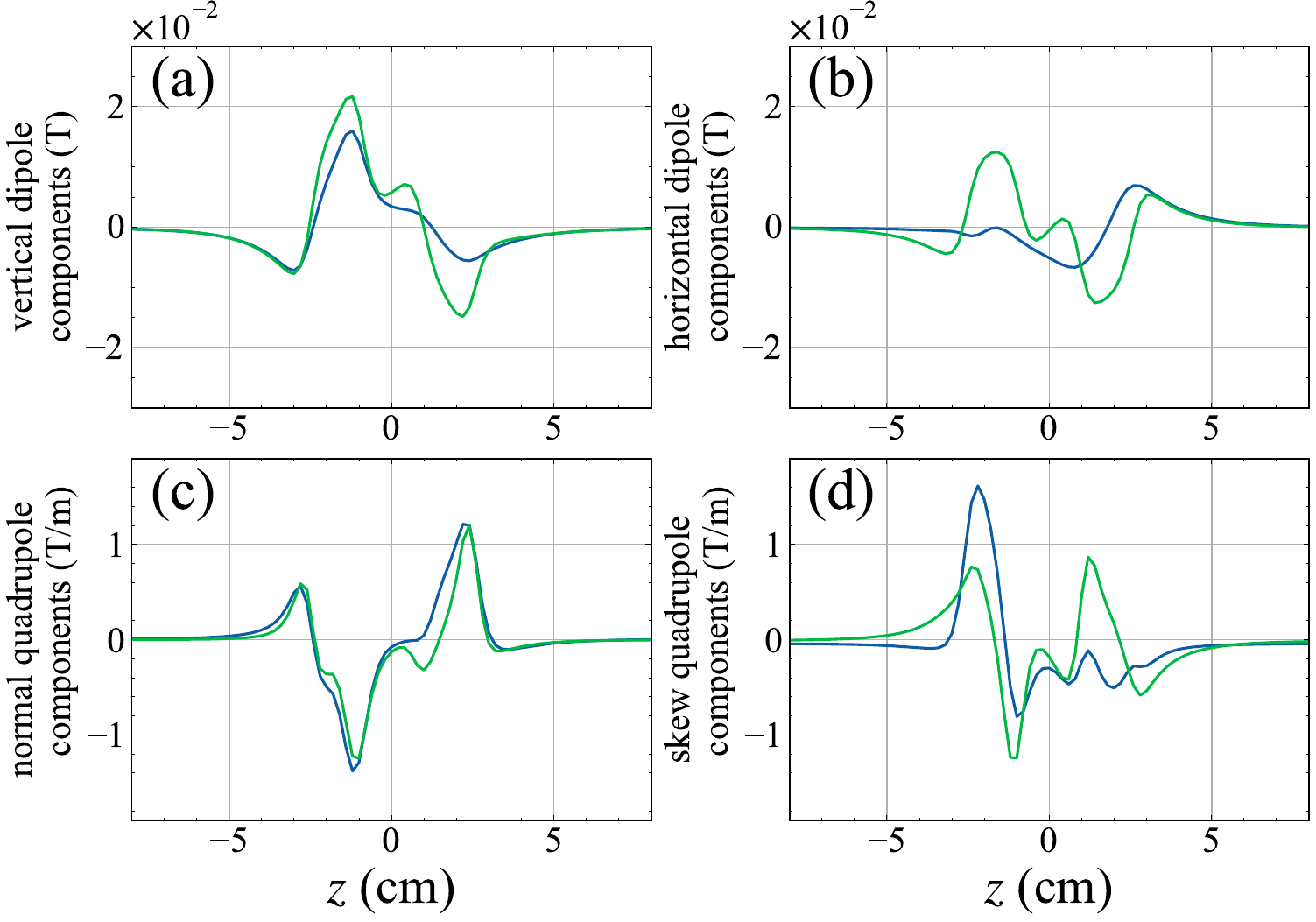}
\caption{(a) Vertical and (b) horizontal dipole components of the measured fields. (c) Normal and (d) skew quadrupole components of the measured fields. Blue traces represent the superposition of single-ring measurements before assembly, while green traces correspond to measurements of the dual ring after assembly.}
\label{fig:pms_multipoles}
\end{figure}

A critical design consideration for PMS systems is the residual steering effect of its stray field when the magnet is retracted from the beam path. Unlike electromagnetic solenoids which can be deactivated, a PMS must be mechanically displaced, potentially exposing the beam to fringe fields when not in use. To quantify this effect, we simulate off-axis stray field maps of the dual RM-PMS ring in {\sc radia}~\cite{Chubar1998} for different horizontal offsets and import field maps into {\sc gpt}~\cite{gpt} for beam tracking. Result in Figure~\ref{fig:pms_steering} shows significant steering effect in the $x$ direction, with induced kick over 15 mrad when moved 5 cm off-axis for an electron beam with 4~MeV kinetic energy. The steering diminishes at larger offsets, suggesting adequate suppression can be achieved by retracting the magnet sufficiently from the beam path. Although magnetic shielding could further reduce the steering~\cite[Chapter 4]{gehrke_pms_2013}, adding it to the 1.4 kg weight of the dual PMS ring would increase the complexity to the stage assembly. Consequently, we have opted to mitigate the steering effect by ensuring a sufficient horizontal displacement rather than incorporating additional shielding.

There are two primary techniques for the manufacturing of RM-PMS rings: hot-pressed sintering~\cite{grunberger1997hot} (producing monolithic rings with superior azimuthal uniformity) and wedge-based assembly (using pre-magnetized sectors). While the former offers enhanced uniformity in magnetization~\cite{xu:ipac24}, the geometric constraints of our design---specifically the ring length and wall thickness---necessitate the wedge-based approach. Field maps derived from {\sc radia} simulation indicate that segmentation of eight pieces does not generate significant multipole components under ideal conditions (though imperfections of magnetization may introduce such components, as discussed in the following section). Notably, the peak on-axis field is reduced by approximately 2.5\% compared to a monolithic RM-PMS ring---a compromise deemed acceptable given the manufacturing feasibility.

\subsection{Characterization}\label{subsec:characterization}

\begin{figure*}[t]
\centering
\includegraphics[width=1.95\columnwidth]{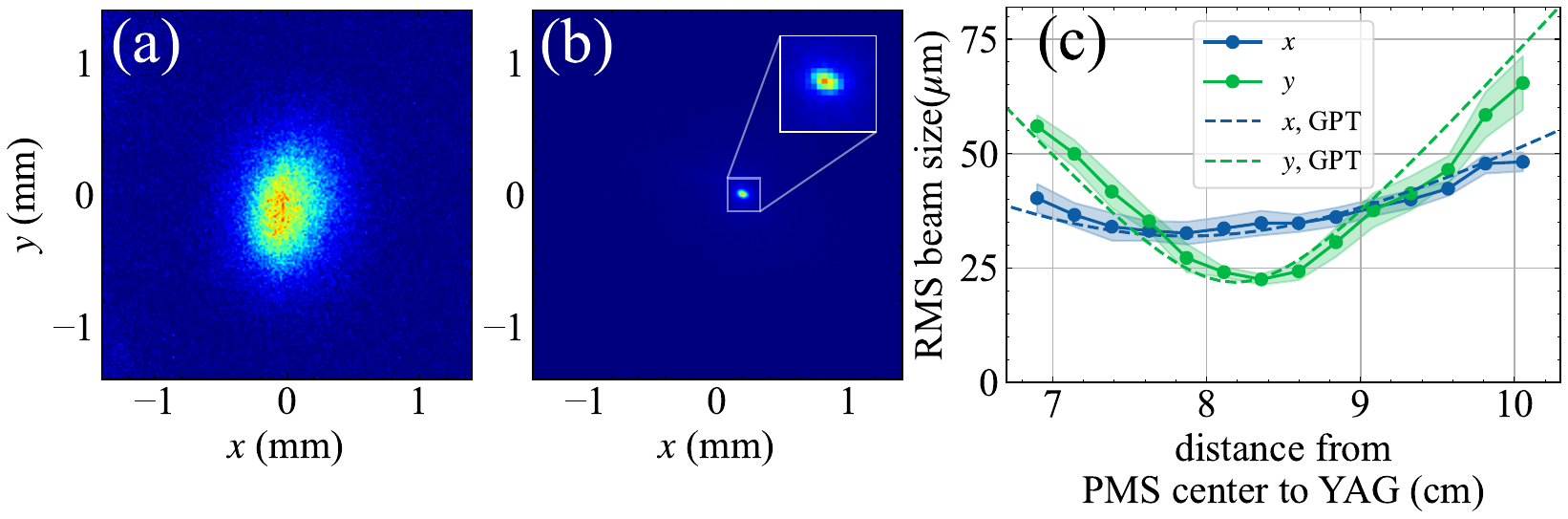}
\caption{Measured electron beam profiles on the YAG screen when PMS is retracted (a) and inserted (b) in beam path. Inset in (b) shows a zoom-in view of the focused electron beam. (c) RMS beam size vs. z position of the PMS in axial scan. The shaded area in blue and green corresponds to one standard deviation around the average beam sizes at each longitudinal position. The standard deviation and average are calculated over approximately 30 shots after applying a threshold on beam intensities.}
\label{fig:pms_beam_profile}
\end{figure*}

The wedge-based fabrication process introduces susceptibility to magnetization imperfections in individual wedges~\cite{hachmann_design_2016}. To mitigate these effects, each single RM-PMS ring underwent systematic characterization, enabling sorting and rotational pairing to minimize the steering and astigmatism from field errors. Upon receiving the single RM-PMS rings from the manufacturer, we implemented the following procedures for characterization and assembly:
\begin{enumerate}
\item \textbf{3D field mapping}: 3D Hall probe scans (0.4 cm$\times$0.4 cm$\times$40 cm volume around mechanical and field centers) were performed for all RM-PMS rings.
\item  \textbf{Generalized gradient computation}: The measured field maps around mechanical centers were processed to compute the generalized gradient expansion of the magnetic fields~\cite{venturini_accurate_1999}. This technique expresses the fields in terms of $z$-dependent multipoles coefficients and their derivatives (pseudo-multipoles) and is derived from the measured $(B_{x},B_{y},B_{z})$ components on the surface of a rectangular boundary. The expansion effectively smooths measurement noise and errors, providing a robust and accurate representation of the field for transfer map calculations. In this work, field maps centered on the mechanical axis were used instead of the magnetic field centers, as the magnets are to be rotated about their mechanical centers during assembly.
\item \textbf{Numerical optimization}: Generalized gradient representation of each RM-PMS is numerically rotated and paired with a counterpart of opposite magnetization for all possible permutations. Transfer matrices of the combined fields were calculated in {\sc elegant}~\cite{borland2000elegant}, optimizing rotation angle and combination to minimize deviations in $R_{21}$ and $R_{43}$ (astigmatism) while suppressing beam kicks induced by dipole components.
\item \textbf{Assembly and validation}: Selected pairs were physically rotated, assembled, followed by a 3D Hall probe scan to verify the field components after combination.
\end{enumerate}

In principle, step 4 can be iterated to refine focusing performance (stigmatic focusing and zero kick) based on measured fields. However, the strong axial attraction forces between oppositely magnetized RM-PMS rings complicate disassembly and angular adjustments. Consequently, the numerical superposition in step 3 served as the definitive optimization step. A batch of ten single RM-PMS rings was ordered from the vendor---five radially magnetized inward and five outward. Generalized gradient characterization confirmed that the peak on-axis fields of all rings are within 0.01~T from the expected values. Based on these results, two pairs of RM-PMS rings were selected and assembled following the outlined procedure. 

Figure~\ref{fig:pms_measurements}(a) shows the picture of a dual RM-PMS ring after assembly. The generalized gradients calculated from 3D field maps contain solenoidal and multipole components. We compare three evaluations of the solenoidal fields $B_{z}$ in Fig.~\ref{fig:pms_measurements}(b): (1) superposition of single ring measurements, (2) measurements of the dual ring after assembly, and (3) fields computed from Eq.~(\ref{eq:Bz_dual_pms}) based on ring dimensions. The excellent agreement between the three methods confirms that the dual RM-PMS ring achieves peak field of 1~T. Similarly, the multipole components of the dual RM-PMS ring are shown in Fig.~\ref{fig:pms_multipoles} (a)-(d). Overall we found reasonable agreement between pre- and post-assembly field measurements, validating the robustness of the characterization and assembly procedure.


\section{Electron Beam Experiment}

To confirm the focusing properties of the dual PMS ring, we performed electron beam experiments at the UCLA Pegasus beamline. The beamline consists of an S-band rf photoinjector comprising an alkali antimonide photocathode mounted in a 1.6-cell resonant cavity. Electron bunches are produced from the gun with 500~fC charge, focused by a gun solenoid, and injected into a downstream linac to accelerate the beam to 7.1~MeV kinetic energy. A dual RM-PMS ring is mounted on a 3-axis translation stage located 3.6~meters downstream of the cathode. This stage enables horizontal translation over a distance of 7~cm, allowing the PMS to be extracted from the beam path when needed. In addition, the longitudinal stage offers a 5~cm travel range, while a vertical actuator permits adjustments of $\pm$1~cm. Transverse beam profiles are recorded using a 20~$\mu$m thick YAG screen located 8.2~cm from the center of the PMS.

Figures~\ref{fig:pms_beam_profile} (a) and (b) present the electron beam profiles with the PMS retracted and inserted. With PMS inserted, the RMS beam size decreases from 215~$\mu$m in $x$ and 315~$\mu$m in $y$ (average of 52 shots), to 35~$\mu$m in $x$ and 25~$\mu$m in $y$ (averge of 30 shots), demonstrating its focusing effect. Note that although the beam profiles appear rotated by $90^{\circ}$ between the two configurations, this is not primarily due to Larmor rotation---the $x$-$y$ coupling introduced by the PMS is relatively small (to be shown in the following)---but is attributed to asymmetric Twiss parameters in the $x$ and $y$ planes.

To accurately determine the beam waist, we performed a longitudinal scan of the PMS position, recording approximately 30 shots of beam profiles at each position. Figure~\ref{fig:pms_beam_profile}(c) illustrates the evolution of the RMS beam size as a function of the axial displacement of the PMS. The beam size reaches a minimum of 34~$\mu$m in $x$ and 23~$\mu$m in $y$ at a longitudinal position of 8.4~cm, thereby confirming the focusing capability predicted from the magnetic field distribution. 

The axial scan also allows us to determine the emittance and Twiss parameters of the incoming electron beam. The RMS beam size evolution is fitted using the function $\sigma(s)$ with $\sigma(s) = \sqrt{\beta_{0}{\varepsilon_{\mathrm{geo}}}\left[1+(\frac{s-s_{0}}{\beta_{0}})^{2}\right]}$ where $\beta_{0}$ is the Twiss $\beta$ function at waist, $s_{0}$ is the waist location, and $\varepsilon_{\mathrm{geo}}$ is the geometric emittance. This allows  us to extract the geometric emittance of the beam $\varepsilon_{\mathrm{geo}}$ immediately upstream of the focus, yielding horizontal and vertical values of 59 nm and 82 nm, respectively. The Twiss parameters at PMS entrance are obtained by numerically matching the beam size evolution in {\sc gpt} simulation with measured values. The initial Twiss parameters upstream of the PMS (25~cm from magnet center) are found to be $\alpha_{x}$=0.6, $\beta_{x}$=0.55~m, $\alpha_{y}$=0.02, $\beta_{y}$=1.2~m, with fitted beam evolution shown in Figure~\ref{fig:pms_beam_profile}(c).


\begin{figure*}[t]
\centering
\includegraphics[width=1.95\columnwidth]{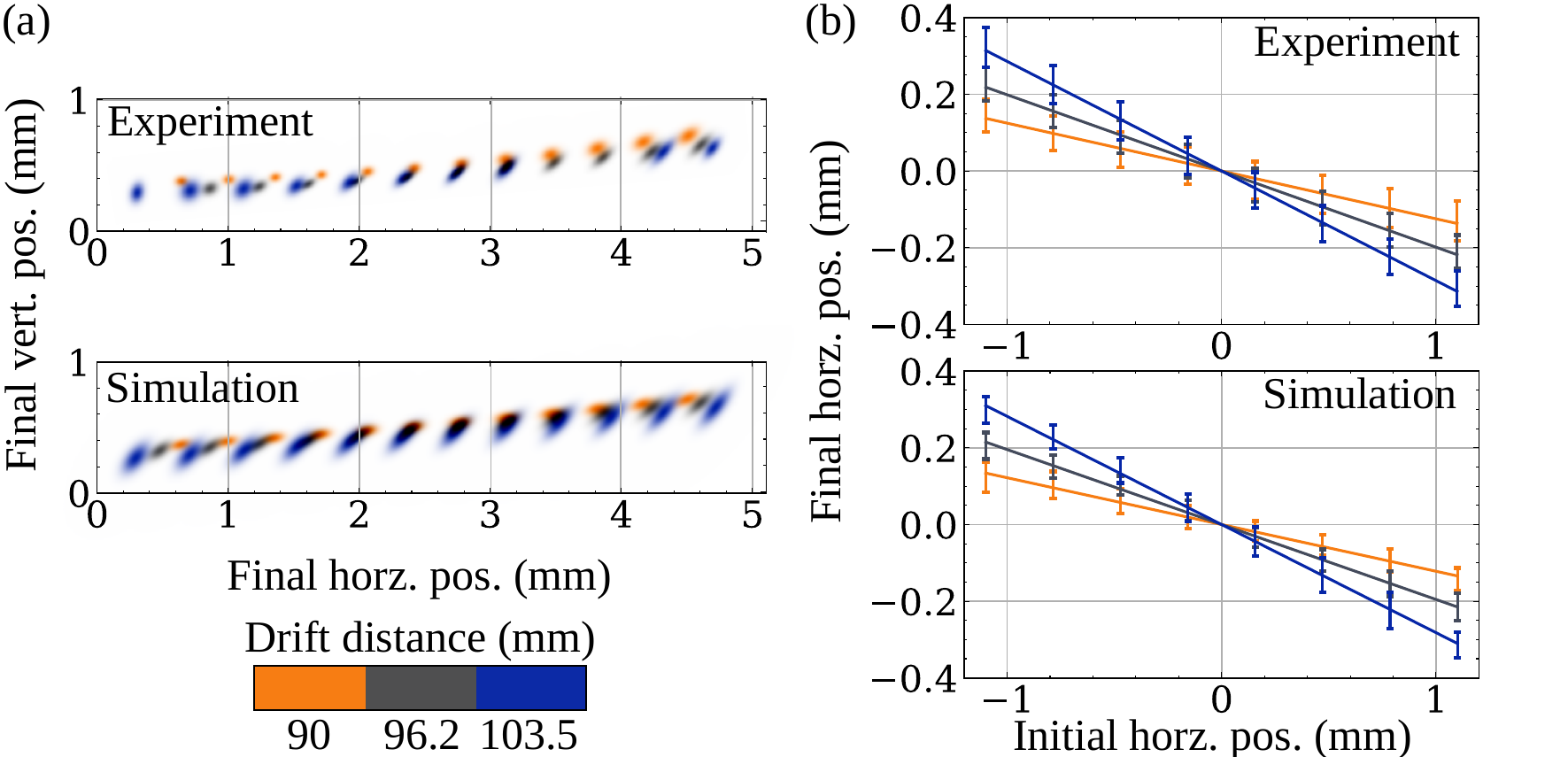}
\caption{Linear transport measurements: (a) horizontal scan of the PMS lens position results in the image on the final screen shown, color encodes the longitudinal distance from magnet center: top panel shows experimental data and bottom panel particle-tracking simulation.  (b) Horizontal position of the spot on the screen as a function of the position of the beam in the lens, both expressed relative to the optical axis of the lens, for three longitudinal lens positions shown in the legend. Top panel shows experimental data, bottom simulation. Error bars show $\pm \sigma$, the RMS size of the beam centroid movement on the screen. Color carries ths same meaning as (a).}
\label{fig:transport}
\end{figure*}

To characterize experimentally the linear transport of the PMS, we scanned the horizontal and longitudinal lens position on the motorized stage, and measured the corresponding changes in the beam centroids on the YAG sceen. A superposition of the measured beam profiles is presented in Figure~\ref{fig:transport}(a). Denoting the horizontal and longitudinal displacements of the stage $X,Z$, and the horizontal and vertical beam centroids at the YAG screen $\langle x_{1} \rangle$ and $\langle y_{1} \rangle$, the following expressions for linear transport matrix elements can be derived:
\begin{align}
R_{11} &= 1-\partial_X \langle x_1\rangle, \label{R11} \\
R_{21} &= \partial_X \partial_Z \langle x_1\rangle,\label{R21} \\
R_{31} &= -\partial_X \langle y_1\rangle, \label{R31} \\
R_{41} &= \partial_X \partial_Z \langle y_1\rangle.\label{R41}
\end{align}
These expressions follow from a coordinate transformation that maps the frame in which the lens is stationary (with respect to which the transport matrix is defined) to the laboratory frame (in which the position of the beam as it enters the lens is held fixed and instead the center of the lens moves). Written explicitly, the change of variables is $x_1 \mapsto x_1 + X$, $\partial_{x_0} \mapsto -\partial_{X}$ and $\frac{d}{dz} \mapsto -\partial_Z$. Partial derivatives in Eqs~\eqref{R11}-\eqref{R41} are extracted from polynomial fits to the discrete scan data:
\begin{align}
\langle x_1 \rangle &= c_0 + c_1 X + c_2 Z + c_3 X Z, \\
\langle y_1 \rangle &= c_4 + c_5 X + c_6 Z + c_7 X Z,
\end{align}
with $c_i$ the fit parameters. The agreement of these fits with data is shown in Fig.~\ref{fig:transport}(b), with coordinates transformed back into the lens-stationary frame. Solid lines show the fitted relationship between horizontal beam position entering the lens and position measured on the screen for three different lens-to-screen drift distances. The trend lines show the beam becoming more defocused with increasing drift distance, and the focal length is determined by the rate at which defocus increases with drift distance [summarizing in words the meaning of Eq.~\eqref{R21}].

\begin{figure*}[t]
\centering
\includegraphics[width=1.95\columnwidth]{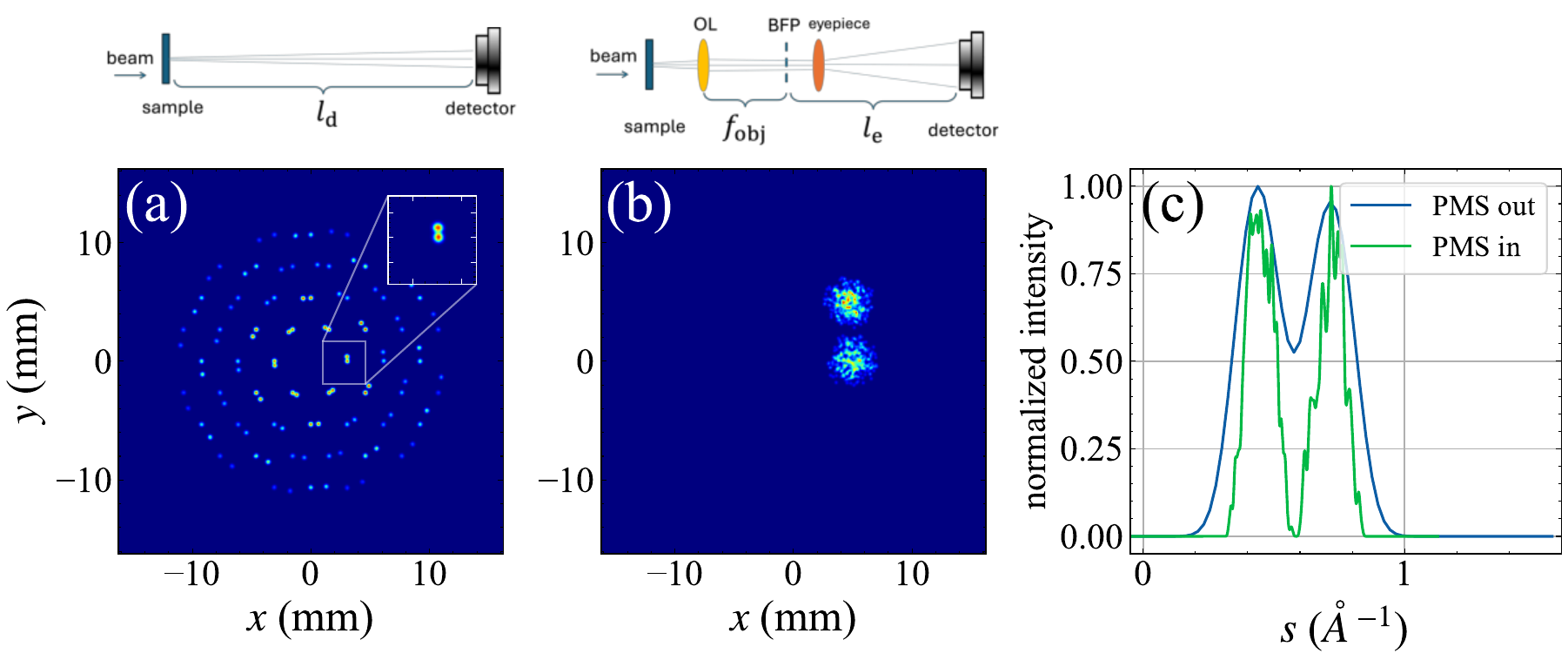}
\caption{Simulated diffraction pattern of bilayer $\mathrm{WS}_{2}$ when PMS magnets are retracted (a) and inserted (b) in the beam path. The inset in (a) shows the region of interest when zoomed in. The cartoons on the top depict the trajectories of the scattered electron for the two cases. (c) projections of two diffraction peaks with coordinate converted to reciprocal space for PMS out (blue trace) and PMS in (green trace). }
\label{fig:pms_angular_magnification}
\end{figure*}

\begin{table}[b]
    \centering
    \begin{tabular}{ccc}
        \hline\hline
         & \textbf{Experiment} & \textbf{ Simulation } \\ \hline
        $R_{11}$ & $-3 \pm 1 \times 10^{-2}$                   & $-3.57\times10^{-2}$\\ 
        $R_{21}$ & $-11.9 \pm 0.3 \ \mathrm{m}^{-1}$               & $-11.9 \ \mathrm{m}^{-1}$ \\ 
        $R_{31}$ & $8 \pm 2 \times 10^{-2}$                   & $8.70 \times 10^{-2}$ \\
        $R_{41}$ & $0.5 \pm 0.5 \ \mathrm{m}^{-1}$ & $7.65 \times 10^{-1} \ \mathrm{m}^{-1}$ \\ \hline \hline
    \end{tabular}
    \caption{Transport matrix elements and fitting errors from magnet entrance to 8.25 cm downtream of magnet center, for a beam kinetic energy of 7.1 MeV.}
    \label{tab:transport_matrix}
\end{table}

We also simulate the horizontal scans in {\sc gpt} using the 3D field maps reconstructed from generalized gradient characterization described in Sec.~\ref{subsec:characterization}, and repeat the fitting procedures for the simulated scan. The results of the fitting are shown in Fig.~\ref{fig:transport} (b), with the extracted transport matrix elements summarized in Table~\ref{tab:transport_matrix} for experiment and simulation. It is worth noting that the fitting uncertainties for  $R_{11}$, $R_{31}$, $R_{41}$ appear relatively large due to the small absolute values of these matrix elements---small variations in the data lead to proportionally larger relative errors. A key matrix element is $R_{21}$, which relates to the focal length $f$ through $f=-1/R_{21}$. The experimentally determined focal length is $f = 8.4 \pm 0.1 \ \mathrm{cm}$, consistent with the simulation predictions. Additionally, small but non-negligible $R_{31}$ and $R_{41}$ terms indicate some degree of $x$-$y$ coupling. The coupling arises from skew quadrupole components introduced by imperfections in the wedge geometry.

\section{Applications}

\subsection{Angular magnification}
Our primary intended application of the dual RM-PMS ring is angular magnification in MeV-UED~\cite{denham_high_2024}. In a typical UED beamline, electrons are scattered by a sample and the scattering angles $x_{\mathrm{sample}}^{\prime}$ are converted into spatial offsets $x_{\mathrm{detector}}$ on the detector by a drift $l_{d}$, 
\begin{equation}
    x_{\mathrm{detector}} = x_{0} + l_{d} (x_{0}^{\prime}  + x_{\mathrm{sample}}^{\prime}) 
\label{eq:mapping_drift}
\end{equation}
here $x_{0}$ and $x_{0}^{\prime}$ are the position and divergence of the electron just before the sample. The mapping from $x_{\mathrm{sample}}^{\prime}$ to $x_{\mathrm{detector}}$ is not exact due to the contribution of $x_{0}$ and $x_{0}^{\prime}$. Adding magnetic optics downstream of the sample can remove the $x_{0}$ term and increase the effective camera length (transfer matrix element $R_{12}$). As illustrated in Fig.~\ref{fig:pms_angular_magnification}, an objective lens (OL) of focal length $f_{\mathrm{obj}}$ maps the divergence into spatial offsets at its back focal plane (BFP). The virtual image is then magnified by an eyepiece lens of focal length $f_{\mathrm{eye}}$ over distance $l_e$. The final position on detector is given by
\begin{equation}
\begin{aligned}  
    x_{\mathrm{detector}} & = f_{\mathrm{obj}} \frac{l_{e} + \sqrt{l_{e}(l_{e} - 4f_{\mathrm{eye}})}}{l_{e} - \sqrt{l_{e}(l_{e} - 4f_{\mathrm{eye}})}}  (x_{0}^{\prime}  + x_{\mathrm{sample}}^{\prime}) \\
    & \approx  f_{\mathrm{obj}} \frac{  l_{e} } {f_{\mathrm{eye}}} (x_{0}^{\prime}  + x_{\mathrm{sample}}^{\prime})
\end{aligned}
\label{eq:mapping_magnification_1}
\end{equation}
the approximation above is valid when $l_{e}\gg f_{\mathrm{eye}}$. With proper choice of of $l_{e}$, $f_{\mathrm{obj}}$, and a short-focal-length eyepiece lens (small $f_{\mathrm{eye}}$), the camera length can be made larger than $l_{d}$ which allows magnification of the diffraction pattern.

Here we consider an angular magnification setup at SLAC MeV-UED facility~\cite{weathersby2015mega}. The objective lens is an axially magnetized AM-PMS with focal length $f_{\mathrm{obj}}=67$~cm at 4~MeV~\cite{ampms-vendor}. The eyepiece lens is the dual RM-PMS ring described in previous sections, with a focal length $f_{\mathrm{eye}}=4.2$~cm. For a total drift distance $l_{d}$=3.2~m, we estimate from Eq.~(\ref{eq:mapping_magnification_1}) that a magnification of 12 can be achieved compared with drift only when objective lens is positioned 10~cm downstream of sample. We perform start-to-end {\sc gpt} simulations for a bilayer $\mathrm{WS_{2}}$ sample with 7$^\circ$ tilt between layers, where the final electron beam $(x,y)$ distributions are binned into images and convolved with the point spread function (PSF) of the phosphor screen (Gaussian with $\sigma$=85~$\mu$m). The diffraction patterns on the detector are shown in Fig.~\ref{fig:pms_angular_magnification} for scenarios when the PMS magnets are retracted and inserted. Without the magnification, the resolution in the reciprocal space suffers from PSF of the screen and finite pixel size. The angular magnification enabled by the post-sample optics overcomes this limitation; the diffraction peaks in Figure~\ref{fig:pms_angular_magnification}(b) are clearly separated, whereas they are connected in the unmagnified case. 

To quantify the reciprocal space resolution, we converted the diffraction patterns into reciprocal space to determine the widths of the diffraction peaks. As shown in Figure~\ref{fig:pms_angular_magnification} (c), the RMS spread of the diffraction peaks (relative to a predefined centroid separation of 0.278 $\AA^{-1}$ for adjacent 100 peaks) is determined by Gaussian fitting of the peak width. This analysis yields a reciprocal resolution of 0.083 $\AA^{-1}$ for the unmagnified case and 0.05 $\AA^{-1}$ for the magnified case. This improvement is consistent with the resolution calculated from the beam size $\sigma_{x}$ and normalized emittance $\varepsilon_{x}$ at the sample plane using $\Delta s=\frac{2\pi}{\lambda_{e}}\frac{\varepsilon_{x}}{\sigma_{x}}$, where $\lambda_{e}$ is the Compton wavelength of electrons~\cite{weathersby2015mega}. Further resolution enhancement will require reducing the beam emittance.

Lens aberrations do not affect the magnified image because the beamlets of interest at the PMS entrance are much smaller than the lens aperture. However, limited by the finite detector size, only a few magnified diffraction peaks can be captured within a single field of view. To address this, a steering magnet can be used to direct different portions of the diffraction pattern through the PMS, enabling targeted magnification of specific regions in reciprocal space. Using PMS for both the objective and the eyepiece lenses offers an additional benefit: it cancels net Larmor rotation, which further simplifies the alignment and navigation across regions of interest.

\subsection{Tight focusing}

A variety of accelerator applications require tight focusing of MeV electron beams, e.g., inverse Compton scattering. Simulation and measurements indicate the PMS should deliver a focal length of a few cms and achieve sub-10 $\mu$m beam sizes for MeV electron beams. In an ideal lens, the RMS beam size at the focus is limited only by the convergence angle $\varphi$ of the beam, a result that follows from the envelope equation: 
\begin{equation}
    \sigma_{x,f} = \frac{\varepsilon_{x}}{\beta\gamma\varphi}.
\label{eq:rms_at_waist_linear}
\end{equation}
Here, $\varepsilon_{x}$ is the normalized emittance, $\beta$ is the ratio of $v$ to $c$, $\gamma$ is the Lorentz factor, and $\varphi := \sigma_{x,0}/f$ with $\sigma_{x,0}$ the beam size at lens entrance. With short-focal-length magnets and low-emittance electron beams, a sub-10$\mu$m beam size can be achieved simply by increasing the initial beam size. 

\begin{figure}[t]
\centering
\includegraphics[width=0.95\columnwidth]{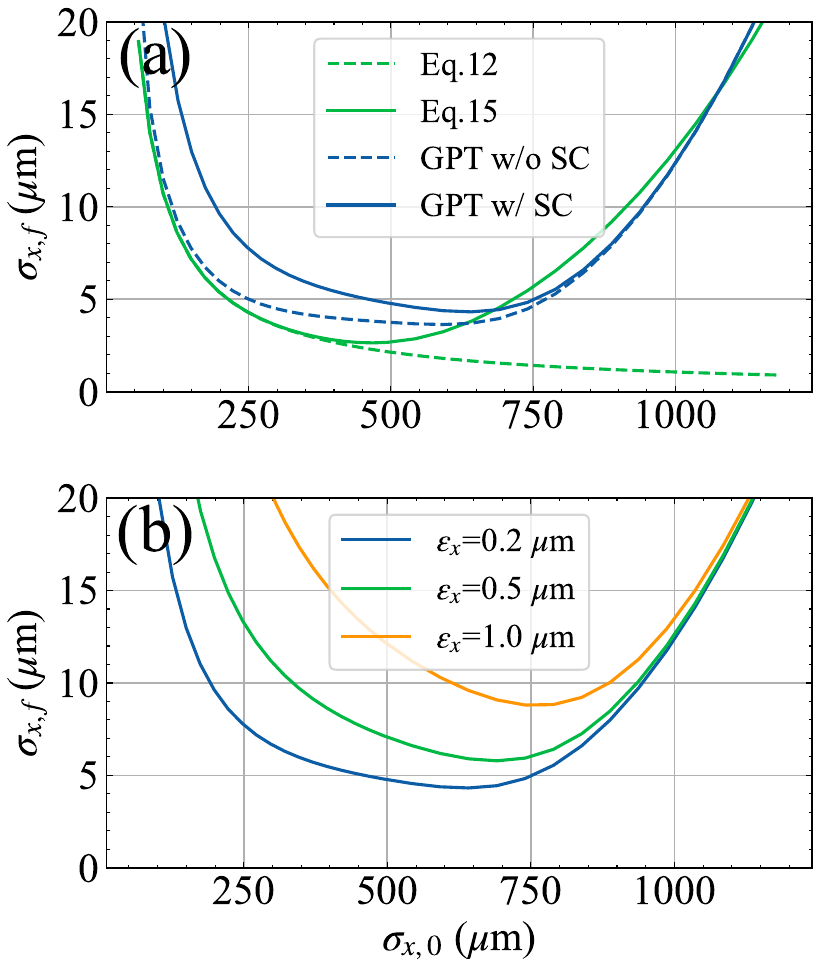}
\caption{Evolution of beam size at focal point vs. beam size at lens entrance. (a) Comparison of waist sizes calculated with different methods for a beam with 0.2~$\mu$m emittance. (b) beam sizss at focal point for beams with different emittances (simulated using {\sc gpt} with space charge effects).}
\label{fig:pms_focusing_x0}
\end{figure}

However, nonlinear aberration of the lens imposes a practical limit on this scheme. A simple transport model can be used to estimate the minimum beam size achievable with the PMS, accounting for the nonlinear focusing of the magnet.
Consider a particle with trace space coordinates $(x_{0},x_{0}^{\prime})$ at the magnet entrance. The particle receives a focusing kick when passing through the magnet,
\begin{equation}
    x_{1}^{\prime} = x_{0}^{\prime} + R_{21} x_{o} + U_{2111}x_{0}^{3}.
\label{eq:kick_at_magnet}
\end{equation}
The matrix element $R_{21}$ can be approximated by $-1/f$, while $U_{2111}$, the higher order term associated with the the nonlinear aberration, can be computed exactly with {\sc cosy infinity} or Green's function methods~\cite{denham2021}. For beams with higher energy, $U_{2111}$ can also be calculated from the field profile following Eq.~(6) in ~\cite{lund:ipac15-thpf139} under a constant-radius approximation. Assuming the transverse coordinate at the magnet exit is unchanged ($x_{1}=x_{0}$), the position of the particle in the back focal plane is:
\begin{equation}
    x_{f}  = x_{1} + f x_{1}^{\prime} =  x_{0} + fx_{1}^{\prime}.
\label{eq:x_f_at_waist}
\end{equation}
Inserting Eq.~(\ref{eq:kick_at_magnet}) into Eq.~(\ref{eq:x_f_at_waist}), the resulting expression for the RMS beam size, valid for a Gaussian beam, is:
\begin{equation}
    \sigma_{x,f}  = f\sqrt{\frac{\varepsilon_{x}^{2}}{\beta^{2}\gamma^{2}\sigma_{x,0}^2} + 15 U_{2111}^{2} \sigma_{x,0}^6}.
\label{eq:rms_at_waist}
\end{equation}
In deriving this expression we make use of the mathematical identity that, for a Gaussian distribution, $\left<x_{0}^{6}\right>=15\sigma_{x,0}^{6} $. The optimal initial size to achieve the minimum final spot size is therefore,
\begin{equation}
    \sigma_{x,0}^{\mathrm{min}} = {\left(\frac{\varepsilon_{x}^{2}}{45\beta^{2}\gamma^{2}U_{2111}^{2}}\right)}^{\frac{1}{8}}.
\end{equation}

We calculated the spot sizes at the focal point of the PMS ring for a 7 MeV electron beam with 500~fC charge and various initial sizes, and compared the results obtained with different methods. The focal length and $U_{2111}$ of the dual RM-PMS ring are calculated to be 8.4~cm and -0.045 $\mathrm{cm}^{-3}$ respectively at kinetic energy of 7~MeV\footnote{The results are calculated with {\sc cosy infinity} and agree with values calculated with Green's function method. Calculating $U_{2111}$ with Eq.~(6) in~\cite{lund:ipac15-thpf139} gives -0.08$\mathrm{cm}^{-3}$, which overestimates the nonlinear kick as it uses a constant-radius approximation. }. As depicted in Fig.~\ref{fig:pms_focusing_x0}(a), the final RMS beam size continues to decrease with larger initial beam size until it reaches a minimum,  beyond which further increases in the initial beam size result in growth that scales like $\sigma_{x,0}^3$ as indicated by Eq.~(\ref{eq:rms_at_waist}). Space charge forces induce a defocusing for the converging beam at the waist location, effectively increasing the beam size at the nominal focal point. In general, the beam sizes follow the trends predicted by Eq.~(\ref{eq:rms_at_waist}). 

We also compare the beam sizes at the back focal plane for different initial emittances in {\sc gpt} simulations that include space charge effects. Due to the interplay between the terms in Eq.~(\ref{eq:rms_at_waist}), the impact of nonlinear aberrations is more pronounced for low-emittance beams, while smaller beam sizes are also achieved with lower-emittance beams. Consequently, identifying the smallest possible spot size requires optimally balancing the initial beam sizes with beam emittance and lens aberration considerations.

\section{Conclusions}
We have designed, characterized, and experimentally tested a compact PMS capable of providing strong focusing for MeV-scale electron beams, with a focal length on the order of a few centimeters. The dual-ring, radially magnetized PMS was optimized to achieve a peak on-axis magnetic field of 1~T within a 1.2~cm bore, offering a practical, power-free alternative to conventional electromagnetic solenoids. Detailed field mapping showed excellent agreement with theoretical models, validating both the design methodology and fabrication quality.

Beam tests conducted using a 7~MeV photoinjector validated the strong focusing capability of the PMS, demonstrating a significant reduction in beam size and confirming its suitability for high-brightness electron beam applications. Complementary simulations further highlighted the PMS’s utility in enabling angular magnification for MeV-UED experiments, improving reciprocal space resolution beyond the intrinsic limits of the detector by overcoming point spread function and pixel size constraints.

Beyond ultrafast electron diffraction, the PMS design is broadly applicable to other accelerator-based systems requiring compact, high-field focusing for moderately relativistic (few MeV) beams. Potential applications include inverse Compton scattering sources, where tight focusing is critical for high brightness, as well as MeV-scale electron microscopy, where high-resolution imaging relies on strong, aberration-minimized magnetic lenses. The demonstrated performance establishes this PMS as a robust solution for precision beam control in compact accelerator systems. Future work will focus on further optimization of its design for enhanced focusing strength and reduced aberrations, as well as exploring its integration into broader accelerator applications.

\ 
\section{Acknowledgements}
We would like to thank Marcos Ruelas (Radiabeam) for suggestions on permanent magnet vendors. This work was supported by the U.S. Department of Energy Office of Science, Office of Basic Energy Sciences under Contract No. DE-AC02-76SF00515. The SLAC MeV-UED program is supported by the U.S. Department of Energy Office of Science, Office of Basic Energy Science under FWP 10075, FWP 100713, and FWP 100940. The UCLA Pegasus laboratory operation was partially supported by DOE grant DE-SC0009914. P. Denham was supported by National Science Foundation under Grant No. DMR-1548924. B. Schaap, D. Garcia and A. Kulkarni were supported by National Science Foundation under Grant No. PHY-1549132.

\bibliographystyle{ieeetr}
\bibliography{sample}

\begin{thebibliography}{10}

\bibitem{Musumeci2018review}
P.~Musumeci, J.~{Giner Navarro}, J.~Rosenzweig, L.~Cultrera, I.~Bazarov, J.~Maxson, S.~Karkare, and H.~Padmore, ``Advances in bright electron sources,'' {\em Nucl. Instrum. Methods Phys. Res., Sect. A}, vol.~907, pp.~209--220, 2018.

\bibitem{Filippetto2022review}
D.~Filippetto, P.~Musumeci, R.~K. Li, B.~J. Siwick, M.~R. Otto, M.~Centurion, and J.~P.~F. Nunes, ``Ultrafast electron diffraction: Visualizing dynamic states of matter,'' {\em Rev. Mod. Phys.}, vol.~94, p.~045004, Dec 2022.

\bibitem{graves2014}
W.~S. Graves, J.~Bessuille, P.~Brown, S.~Carbajo, V.~Dolgashev, K.-H. Hong, E.~Ihloff, B.~Khaykovich, H.~Lin, K.~Murari, E.~A. Nanni, G.~Resta, S.~Tantawi, L.~E. Zapata, F.~X. K\"artner, and D.~E. Moncton, ``Compact x-ray source based on burst-mode inverse compton scattering at 100 khz,'' {\em Phys. Rev. ST Accel. Beams}, vol.~17, p.~120701, Dec 2014.

\bibitem{england2014}
R.~J. England, R.~J. Noble, K.~Bane, D.~H. Dowell, C.-K. Ng, J.~E. Spencer, S.~Tantawi, Z.~Wu, R.~L. Byer, E.~Peralta, K.~Soong, C.-M. Chang, B.~Montazeri, S.~J. Wolf, B.~Cowan, J.~Dawson, W.~Gai, P.~Hommelhoff, Y.-C. Huang, C.~Jing, C.~McGuinness, R.~B. Palmer, B.~Naranjo, J.~Rosenzweig, G.~Travish, A.~Mizrahi, L.~Schachter, C.~Sears, G.~R. Werner, and R.~B. Yoder, ``Dielectric laser accelerators,'' {\em Rev. Mod. Phys.}, vol.~86, pp.~1337--1389, Dec 2014.

\bibitem{zhang2018segmented}
D.~Zhang, A.~Fallahi, M.~Hemmer, X.~Wu, M.~Fakhari, Y.~Hua, H.~Cankaya, A.-L. Calendron, L.~E. Zapata, N.~H. Matlis, {\em et~al.}, ``Segmented terahertz electron accelerator and manipulator (steam),'' {\em Nature Photonics}, vol.~12, no.~6, pp.~336--342, 2018.

\bibitem{rkli2014}
R.~K. Li and P.~Musumeci, ``Single-shot mev transmission electron microscopy with picosecond temporal resolution,'' {\em Phys. Rev. Appl.}, vol.~2, p.~024003, Aug 2014.

\bibitem{zhou2019demonstration}
Z.~Zhou, Y.~Fang, H.~Chen, Y.~Wu, Y.~Du, L.~Yan, C.~Tang, and W.~Huang, ``Demonstration of single-shot high-quality cascaded high-energy-electron radiography using compact imaging lenses based on permanent-magnet quadrupoles,'' {\em Phys. Rev. Appl.}, vol.~11, no.~3, p.~034068, 2019.

\bibitem{denham_high_2024}
P.~Denham, Y.~Yang, V.~Guo, A.~Fisher, X.~Shen, T.~Xu, R.~J. England, R.~K. Li, and P.~Musumeci, ``High energy electron diffraction instrument with tunable camera length,'' {\em Structural Dynamics}, vol.~11, p.~024302, Mar. 2024.

\bibitem{lim_adjustable_2005}
J.~K. Lim, P.~Frigola, G.~Travish, J.~B. Rosenzweig, S.~G. Anderson, W.~J. Brown, J.~S. Jacob, C.~L. Robbins, and A.~M. Tremaine, ``Adjustable, short focal length permanent-magnet quadrupole based electron beam final focus system,'' {\em Phys. Rev. ST Accel. Beams}, vol.~8, p.~072401, July 2005.

\bibitem{cesar_demonstration_2016}
D.~Cesar, J.~Maxson, P.~Musumeci, Y.~Sun, J.~Harrison, P.~Frigola, F.~O’Shea, H.~To, D.~Alesini, and R.~Li, ``Demonstration of {Single}-{Shot} {Picosecond} {Time}-{Resolved} {MeV} {Electron} {Imaging} {Using} a {Compact} {Permanent} {Magnet} {Quadrupole} {Based} {Lens},'' {\em Phys. Rev. Lett.}, vol.~117, p.~024801, July 2016.

\bibitem{ghaith2019permanent}
A.~Ghaith, D.~Oumbarek, C.~Kit{\'e}gi, M.~Vall{\'e}au, F.~Marteau, and M.-E. Couprie, ``Permanent magnet-based quadrupoles for plasma acceleration sources,'' {\em Instruments}, vol.~3, no.~2, p.~27, 2019.

\bibitem{wan2018design}
W.~Wan, F.-R. Chen, and Y.~Zhu, ``Design of compact ultrafast microscopes for single-and multi-shot imaging with mev electrons,'' {\em Ultramicroscopy}, vol.~194, pp.~143--153, 2018.

\bibitem{hoff2012wide}
B.~Hoff, C.~Chen, J.~Horwath, M.~Haworth, P.~Mardahl, and S.~Heidger, ``Wide aperture permanent magnet solenoid,'' {\em Journal of Applied Physics}, vol.~111, no.~7, 2012.

\bibitem{gehrke_pms_2013}
T.~Gehrke, ``Design of permanent magnetic solenoids for {REGAE},'' Master's thesis, 2013.

\bibitem{hachmann_design_2016}
M.~Hachmann, K.~Flöttmann, T.~Gehrke, and F.~Mayet, ``Design and characterization of permanent magnetic solenoids for {REGAE},'' {\em Nucl. Instrum. Methods Phys. Res., Sect. A}, vol.~829, pp.~270--273, Sept. 2016.

\bibitem{lefranc1982superconducting}
G.~Lefranc, E.~Knapek, and I.~Dietrich, ``Superconducting lens design,'' {\em Ultramicroscopy}, vol.~10, no.~1-2, pp.~111--123, 1982.

\bibitem{Deitrick2018}
K.~E. Deitrick, G.~A. Krafft, B.~Terzi\ifmmode~\acute{c}\else \'{c}\fi{}, and J.~R. Delayen, ``High-brilliance, high-flux compact inverse compton light source,'' {\em Phys. Rev. Accel. Beams}, vol.~21, p.~080703, Aug 2018.

\bibitem{ji2019ultrafast}
F.~Ji, D.~B. Durham, A.~M. Minor, P.~Musumeci, J.~G. Navarro, and D.~Filippetto, ``Ultrafast relativistic electron nanoprobes,'' {\em Communications Physics}, vol.~2, no.~1, p.~54, 2019.

\bibitem{kulkarni24progress}
A.~Kulkarni and P.~Musumeci, ``Progress on pulsed electron beams for radiation effects characterization of electronics,'' {\em Proc. IPAC'24}, pp.~3688--3691, 2024.

\bibitem{peng_axial_2004}
Q.~Peng, S.~McMurry, and J.~Coey, ``Axial magnetic field produced by axially and radially magnetized permanent rings,'' {\em Journal of Magnetism and Magnetic Materials}, vol.~268, pp.~165--169, Jan. 2004.

\bibitem{berz1990}
M.~Berz, ``Computational aspects of optics design and simulation: Cosy infinity,'' {\em Nuclear Instruments and Methods in Physics Research Section A: Accelerators, Spectrometers, Detectors and Associated Equipment}, vol.~298, no.~1, pp.~473--479, 1990.

\bibitem{brown1982}
K.~Brown, ``A {First}- and {Second}-{Order} {Matrix} {Theory} for the {Design} of {Beam} {Transport} {Systems} and {Charged} {Particle} {Spectrometers},'' Tech. Rep. SLAC-r-075, Stanford Linear Accelerator Center, June 1982.

\bibitem{Chubar1998}
O.~Chubar, P.~Elleaume, and J.~Chavanne, ``{A three-dimensional magnetostatics computer code for insertion devices},'' {\em Journal of Synchrotron Radiation}, vol.~5, pp.~481--484, May 1998.

\bibitem{gpt}
``{General Particle Tracer}.'' \url{https://www.pulsar.nl/gpt/index.html}.

\bibitem{grunberger1997hot}
W.~Gr{\"u}nberger, D.~Hinz, A.~Kirchner, K.-H. M{\"u}ller, and L.~Schultz, ``Hot deformation of nanocrystalline nd-fe-b alloys,'' {\em Journal of Alloys and Compounds}, vol.~257, no.~1-2, pp.~293--301, 1997.

\bibitem{xu:ipac24}
T.~Xu, S.~D. Anderson, and R.~J. England, ``{Field characterization of axially and radially magnetized neodymium rings},'' in {\em Proc. IPAC'24}, 2024.

\bibitem{venturini_accurate_1999}
M.~Venturini and A.~J. Dragt, ``Accurate computation of transfer maps from magnetic field data,'' {\em Nucl. Instrum. Methods Phys. Res., Sect. A}, vol.~437, pp.~387--392, 1999.

\bibitem{borland2000elegant}
M.~Borland, ``Elegant: A flexible sdds-compliant code for accelerator simulation,'' tech. rep., Argonne National Lab., IL (US), 2000.

\bibitem{weathersby2015mega}
S.~Weathersby, G.~Brown, M.~Centurion, T.~Chase, R.~Coffee, J.~Corbett, J.~Eichner, J.~Frisch, A.~Fry, M.~G{\"u}hr, {\em et~al.}, ``Mega-electron-volt ultrafast electron diffraction at slac national accelerator laboratory,'' {\em Review of Scientific Instruments}, vol.~86, no.~7, 2015.

\bibitem{ampms-vendor}
``https://www.magnet4less.com/rare-earth-magnets-n45-3-4-in-od-x-1-4-in-id-x-1-4-in-neodymium-ring.''
\newblock product webpage of AM-PMS.

\bibitem{denham2021}
P.~Denham and P.~Musumeci, ``Space-charge aberrations in single-shot time-resolved transmission electron microscopy,'' {\em Phys. Rev. Appl.}, vol.~15, p.~024050, Feb 2021.

\bibitem{lund:ipac15-thpf139}
S.~M. Lund, ``{Nonlinear Optics of Solenoid Magnets},'' in {\em Proc. IPAC'15}, pp.~4048--4050, JACoW Publishing, Geneva, Switzerland.

\end{thebibliography}
\end{document}